# Analyzing Galaxy Catalogues with Minkowski Functionals

Martin Kerscher, Jens Schmalzing & Thomas Buchert

*Theoretische Physik, Ludwig-Maximilians-Universität, Theresienstr. 37, D–80333 München, F.R.G.*

**Abstract.** We discuss the application of Minkowski functionals to galaxy catalogues with emphasis on the boundary corrections and point out the relevance for other statistical methods. The analysis of a cubic subsample of the CfA–1 data serves as an example.

## 1. Introduction

Nowadays various statistical methods have been advanced which attempt to characterize higher–order moments of the density field or higher–order correlations in a given point process, and thus aim to remedy the well–known situation that low–order statistical methods like the two–point correlation function, or the power spectrum, respectively, extract very limited information from density fields or spatial distributions of galaxies. If one characterizes the density field by its power spectrum one gains information on the amplitudes of the density field on different scales, but not on how the different phases of the Fourier components are correlated. Only information on the latter covers the complex morphology of patterns, which in turn enables us to conduct a detailed comparison of theoretical models and observations.

Going beyond the two–point correlation function requires measures which embody, in a well–defined way, informations on *every* order of the hierarchy of correlation functions. It would not be satisfactory if only partial informations of higher–order moments are captured. Also, invariance properties have to be respected: e.g., the measures should neither depend on the orientation in space (rotations), nor on translations, i.e., they should be *motion invariant*. Minkowski functionals, as employed here, have the advantage of satisfying these requirements; moreover they allow for a *complete* characterization of morphology. Some nonlinear statistics, like the void probability function (White 1979) and the genus statistics (see Melott 1990 for a review) can be viewed as members of the family of Minkowski functionals. In particular, as is our concern here, the boundary contribution of a given sample geometry can be calculated for all the functionals separately. However, the deconvolution of the sample geometry is only possible using the complete set of Minkowski functionals, as we shall demonstrate.

Let us now summarize how we proceed. Consider the set of points in three–dimensional space supplied by galaxy coordinates of a catalogue. Let us decorate each point with a ball of radius $r$. Our task is to measure the size, shape and connectivity of the spatial pattern formed by the union set of these balls.

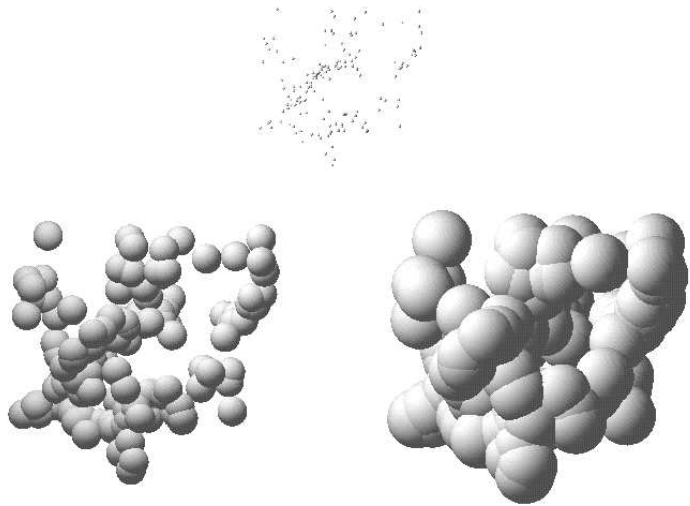

Figure 1.   153 galaxy coordinates from the CfA–1 (according to Gott et al. 1986) decorated with balls of different radius.

These characteristics change with the radius $r$, which may be employed as a diagnostic parameter as illustrated in Figure 1. For the geometric and topological characterization of this union set we use Minkowski functionals (Mecke et al. 1994). Minkowski functionals give us a complete and unique characterization of global morphology (Hadwiger 1957). In three–dimensional space there are four Minkowski functionals having a direct relation to morphological quantities as shown in Table 1. For further details on Minkowski functionals in the cos-mological context see Mecke et al. (1994), Buchert (1995), Platzöder & Buchert (1995) and Schmalzing et al. (1995).

Table 1.   The most common notations for Minkowski functionals in three–dimensional space expressed in terms of the corresponding geometric quantities.

| | geometric quantity | $\mu$ | $M_\mu$ | $V_\mu$ | $W_\mu$ |
|---|---|---|---|---|---|
| $V$ | volume | 0 | $V$ | $V$ | $V$ |
| $A$ | surface | 1 | $A/8$ | $A/6$ | $A/3$ |
| $H$ | integral mean curvature | 2 | $H/2\pi^2$ | $H/3\pi$ | $H/3$ |
| $\chi$ | Euler characteristic | 3 | $3\chi/4\pi$ | $\chi$ | $4\pi\chi/3$ |

## 2. Boundary correction

Integral geometry offers a concise way of dealing with boundaries (Mecke & Wagner 1991). Let $D$ be the window (the sample geometry) through which we look at $N$ galaxies. $B_r = \bigcup_{i=0}^{N} B_r(i)$ is the union of all balls $B_r(i)$ of radius $r$ centered on the $i$–th galaxy respectively. One has to calculate the Minkowski functionals of the intersection of the union of all balls with the window, $M_\mu(B_r \cap D)$ and the Minkowski functionals of the window itself $M_\mu(D)$. Therefore, we have to compute the intersections of balls (Mecke et al. 1994), and also the intersections of balls with the boundary of $D$.

The quantities $M_\mu(B_r \cap D)$ are well suited for the analysis of redshift catalogues since we do not assume anything on the point distribution and still analyze all galaxies in the sample. Using the same window $D$ one is able to compare different catalogues and investigate their homogeneity and isotropy. It is not necessary to impose artificial assumptions on the underlying point process like using periodic boundary conditions (in the case of a cubic sample) or any other way of embedding (e.g. Poisson or vacuum).

Although we argued in favour of *not* imposing assumptions on the underlying point process there is a way of extracting the volume densities of the Minkowski functionals $m_\mu(B_r)$ from the catalogue, assuming homogeneity and isotropy, and thus removing the boundary contribution by deconvolving the window (Schmalzing et al. 1995):

$$m_\mu(B_r) = \frac{M_\mu(B_r \cap D)}{M_0(D)} - \sum_{\nu=0}^{\mu-1} \binom{\mu}{\nu} m_\nu(B_r) \frac{M_{\mu-\nu}(D)}{M_0(D)} \ .$$

Fava & Santaló (1979) give a mathematically rigorous derivation of this formula. In the case of periodic boundary conditions we have $M_\nu(D) = 0$ for $\nu \in \{1, 2, 3\}$ and therefore $m_\mu(B_r) = M_\mu(B_r \cap D)/M_0(D)$ for $\mu \in \{0, 1, 2, 3\}$.
Generally for $\mu = 0$ the formula is trivial. We get for the volume density $m_0(B_r)$ (which is related to the void probability function $P_0(r)$):

$$1 - P_0(r) = m_0(B_r) = \frac{M_0(B_r \cap D)}{M_0(D)} \ .$$

This is the theoretical justification for the Monte Carlo volume integration used by Maurogordato and Lachièze–Rey (1987) for the determination of the void probability function. In a statistical approach (cf. Baddeley 1993 and references therein) one can also prove that this is an unbiased estimator of the void probability.
We want to emphasize that for the removal of boundary contributions of the Euler characteristic in a spatially limited domain $D$ (with non-periodic boundaries) one needs all Minkowski functionals $M_\mu(B_r \cap D)$ and $M_\mu(D)$ with $\mu \in \{0, 1, 2, 3\}$.

## 3. An example

To illustrate this procedure we calculate the density of the Minkowski functionals in the Universe based on the intersection of the Universe with a cube selected from the CfA–1 catalog as discussed by Gott et al. (1986), with side length $s_0 = 5000/\sqrt{3}h^{-1}\text{kms}^{-1}$. In Figure 2 we show the Minkowski functionals $V_\mu(B_r \cap D)$ including the contributions of the intersections with the window domain $D$. The errors are bootstrap errors, we drew ten samples with 143 out of 153 galaxies.

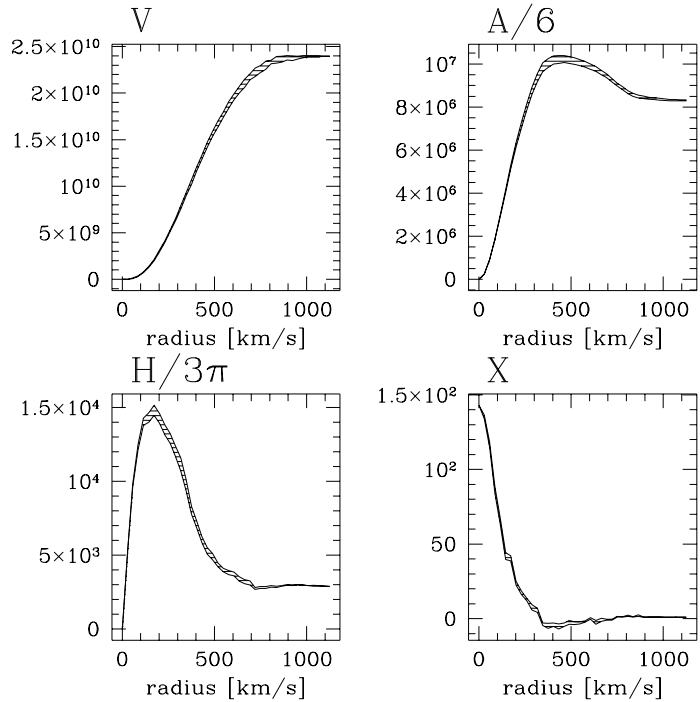

Figure 2. Minkowski functionals of the CfA–1 cube including the window contribution.

Perhaps the most striking feature is the convergence to a constant value for large radii. This is due to the fact that the cube becomes filled up and the functionals $V_\mu(B_r \cap D)$ converge towards the functionals of the cube $V_\mu(D)$. Let us consider this in more detail. The volume is increasing monotonically as expected; the volume density of the spatial domain which is covered by the union set of balls is equal to one minus the void probability of finding a galaxy within that domain. The surface area has a maximum; this is due to the granular structure of the union set for intermediate radii. The same argument applies for the integral mean curvature. For small radii the Euler characteristic has the value 143, since all balls are separated and each ball gives a contribution of unity. As the radius increases, more and more balls overlap and the Euler characteristic decreases. For a radius of $\approx 350 \text{kms}^{-1}$ the Euler characteristic becomes negative due to tunnels in the union set of balls. There is no significant positive maximum after the negative minimum, indicating that only a few cavities form. This suggests a support of less than three dimensions for the distribution of galaxies. Finally, the Euler characteristic converges to unity, which is its value for a cube.

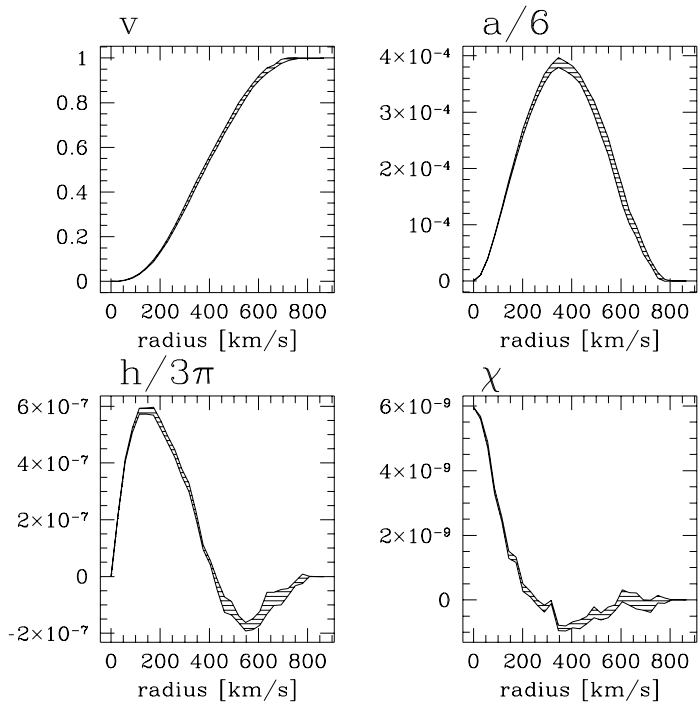

Figure 3. Densities of Minkowski functionals in the Universe based on the CfA–1 cube (after the deconvolution of the window).

In Figure 3 we show the reconstructed densities of the Minkowski functionals $v_\mu(B_r)$ of the Universe, i.e., the density of the Minkowski functionals after removing the contributions of the window $D$. The volume density $v$ converges to 1, the other functionals to 0, showing the already discussed features. The zero of the integral mean curvature density at $\approx 400 \mathrm{kms}^{-1}$ indicates the turnover from a configuration of a connected pattern $B_r$ in a mainly empty space to the configuration consisting of holes through a nearly space filling pattern $B_r$. For the calculation we put, for each radius $r$, a cube with sidelength $s = s_0 - 2r$ (our window $D$) into the center of the sample cube. This is necessary since we have to take into account intersections from balls situated outside the window. For the window we get

$$V_0(D) = s^3, \quad V_1(D) = s^2, \quad V_2(D) = s, \quad V_3(D) = 1.$$

We apply the deconvolution for each radius separately. Again we want to emphasize that *no* assumptions enter in calculating $V_\mu(B_r \cap D)$, but the reconstruction of $v_\mu(B_r)$ is relying on homogeneity and isotropy.

This is meant as an example, this cube certainly not a fair sample.

**Acknowledgments.** We want to thank Herbert Wagner, Vicent Martínez and Maria Jesus Pons for valuable discussions. All authors acknowledge financial support by the "Sonderforschungsbereich 375 für Astro–Teilchenphysik der Deutschen Forschungsgemeinschaft" and enjoyed the hospitality in Valencia. The working visit in Valencia has been supported by "acciones integradas" project AI95–14 (TB) and the "Studienstiftung des deutschen Volkes" (JS).